\documentclass[prd,preprint,tightenlines,floatfix,preprintnumbers,nofootinbib,eqsecnum]{revtex4}

\pdfoutput=1

 \usepackage[dvips,final]{graphicx}
  \usepackage{amssymb}
   \usepackage{amsmath}
    \usepackage{amsfonts}
     \usepackage{epsfig}
      \usepackage{bm} 
      \usepackage[dvipsnames,usenames]{color}
      \usepackage{comment}
      \usepackage{float}
\usepackage[english]{babel}

      \usepackage[utf8]{inputenc}

\usepackage[absolute,overlay]{textpos} 

\begin{document}

\begin{textblock*}{3cm}(17.7cm,.3cm) 
    \noindent\includegraphics[width=3cm]{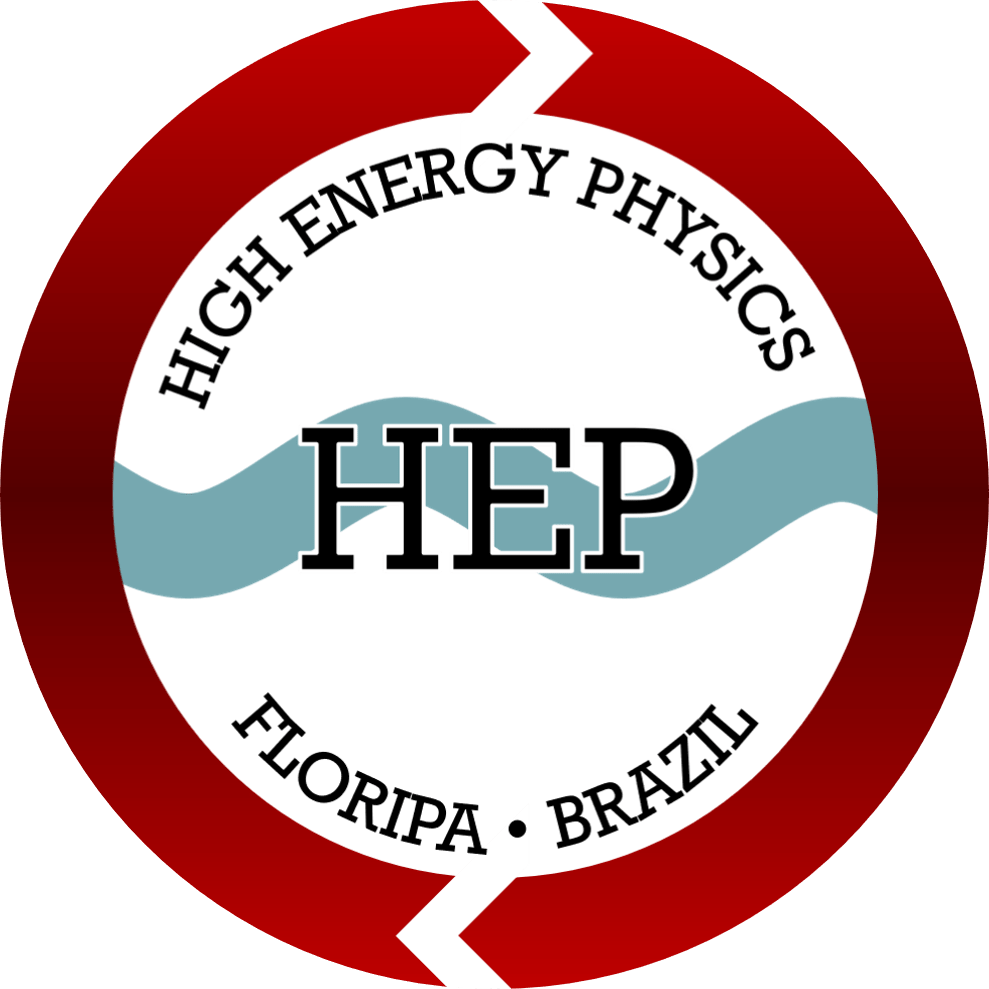} 
\end{textblock*}

\title{
Momentum fraction and hard scale dependence \\ 
of double parton scattering
}

\author{João Vitor C. Lovato}
\email{joaovitorcl1000@gmail.com}

\author{Edgar Huayra}
\email{yuberth022@gmail.com}

\author{Emmanuel G. de Oliveira}
\email{emmanuel.de.oliveira@ufsc.br}

\affiliation{
\\
{$^1$\sl Departamento de F\'isica, CFM, Universidade Federal de Santa Catarina,
C.P. 5064, CEP 88035-972, Florian\'opolis, SC, Brazil}
}


\begin{abstract}
\vspace{0.5cm}

The effective cross section of double parton scattering in high-energy hadron collisions has been measured in proton--proton collisions, with significant variation among final-state observables, contrary to the idea of a universal value. Building upon our previous work, we incorporate the dependence on both the parton longitudinal momentum fraction $x$ and the process energy hard scale $\mu$ into the transverse part of the double parton distributions, using a Gaussian profile. Employing the experimental data from the LHC and Tevatron experiments (covering different processes, kinematic configurations, and center--of--mass energies), we perform a global fit of the model, extracting the parameters that describe the proton structure. With this result, it becomes possible to calculate the effective cross section for other observables, and we provide predictions for future measurements at the LHC.

\end{abstract}

\maketitle

\section{Introduction}
\label{Sec:intro}




In high energy proton proton inelastic collisions, the parton (quarks and gluons) constituents of the proton interact in a complex manner, revealing valuable information about the proton internal structure. Among the various mechanisms involved in these collisions, the best known is single parton scattering (SPS), in which a single pair of partons undergoes a hard interaction and produces the main observable.
As the center--of--mass (c.o.m.) energy increases, so does the number of available partons to participate in the scattering. This leads to a higher probability that more than one hard interaction occurs in a single inelastic event. This phenomenon is known as multiple parton interactions~\cite{Paver:1982yp, Mekhfi:1983az, Sjostrand:1987su}.

The simplest case of multiple parton interactions (MPI) is known as double parton scattering (DPS), in which two partons from each proton initiate two independent hard scatterings within the same proton--proton collision~\cite{Diehl:2011yj, Blok:2011bu, Blok:2013bpa, Diehl:2017wew}. There are various reasons for investigating DPS cross sections at hadron colliders~\cite{Bartalini:2011jp}. To begin with, hadron collisions that involve more than one hard parton scattering can contribute to multi--particle final states~\cite{ATLAS:2018zbr, Kulesza:1999zh}. Also, they can give rise to significant backgrounds to certain rare SPS signals, including new physics~\cite{DelFabbro:1999tf, Hussein:2006xr}. Furthermore, they are an interesting signal in their own right, as they provide new information about the proton structure, e.g., the double parton distribution functions (DPDFs) \cite{Diehl:2023jje, Chang:2012nw, Gaunt:2009re, Diehl:2017kgu, Snigirev:2003cq, Snigirev:2010tk}.

In the above references, if all correlations between partons in the pair are neglected, the cross section of DPS is succinctly encapsulated by the pocket formula:
\begin{align}
\sigma^{\text{DPS}}(AB) = \frac{m}{2} \frac{\sigma^{\text{SPS}}(A)\sigma^{\text{SPS}}(B)}{\sigma_{\text{eff}}}.
\label{eq:pocketformula}
\end{align}
Here, $\sigma^{\text{SPS}}$ represents the standard SPS cross section for the final states $A$ and $B$. The constant $m$ is a symmetry factor: $m = 1$ if $A$ and $B$ are indistinguishable processes, and $m = 2$ otherwise. Finally, the key DPS quantity in Eq.~\ref{eq:pocketformula} is the effective cross section, $\sigma_{\text{eff}}$, which encodes information about the transverse structure of the proton. 

Specific models for the proton have been used in order to calculate the effective cross section~\cite{Calucci:1999yz, DelFabbro:2000ds, Domdey:2009bg, Blok:2010ge, Rinaldi:2013vpa, Huayra:2023gio, Rinaldi:2018slz, Rinaldi:2018bsf, Blok:2022ywz}. Several collaborations have measured the effective cross section or found a limit for it, for different final states, through the measurement of SPS and DPS cross sections~\cite{AxialFieldSpectrometer:1986dfj, Alitti1991ASO, CDF:1993sbj, CDF:1997yfa, ATLAS:2013aph, D0:2014owy, D0:2014vql, CMS:2013huw, D0:2015rpo, CMS:2015wcf, ATLAS:2016rnd, LHCb:2015wvu, ATLAS:2016ydt, LHCb:2016wuo, Lansberg:2017chq, Lansberg:2019adr, CMS:2019jcb, CMS:2021lxi, CMS:2022pio, ALICE:2023lsn, Leontsinis:2022cyi, LHCb:2023qgu, LHCb:2023ybt, CMS:2025xlt, Lansberg:2016muq, Shao:2016wor}. However, the naive use of the formula in Eq.~\ref{eq:pocketformula}  implies the universality of the parameter $\sigma_{\text{eff}}$, which is contrary to observations. Indeed, the available experimental results exhibit considerable variation, with values of $\sigma_{\text{eff}}$ ranging from around 2.7\,mb up to more than 26\,mb, depending on the specific process, c.o.m. energy, and kinematical cuts. 

We would like to have an effective cross section that is dependent on the final state. Building on our previous work~\cite{Huayra:2023gio}, we extend the analysis to incorporate an explicit dependence on both the longitudinal momentum fractions $x$ and the energy hard scale $\mu$ in the transverse double parton distributions, as these two quantities depend on the choice of the final state. We adopt a phenomenological model for the hadron in which the double parton distribution is assumed to follow a Gaussian shape in transverse space, with a variance that depends on $x$ and $\mu$. 

Our model is then used in a global fit to the effective cross sections measured~\cite{CDF:1993sbj, CDF:1997yfa, ATLAS:2013aph, D0:2014owy, D0:2014vql, CMS:2013huw, D0:2015rpo, CMS:2015wcf, ATLAS:2016rnd, LHCb:2015wvu, ATLAS:2016ydt, LHCb:2016wuo, Lansberg:2017chq, Lansberg:2019adr, CMS:2019jcb, CMS:2021lxi, CMS:2022pio, ALICE:2023lsn, Leontsinis:2022cyi, LHCb:2023qgu, LHCb:2023ybt} in several (anti-)proton collision processes involving different final states at the LHC and Tevatron experiments. The quality of the fit surpasses that of our previous study, highlighting the relevance of incorporating $x$- and $\mu$-dependence in the determination of $\sigma_{\text{eff}}$. Moreover, we are able to provide predictions from for several central and forward rapidity final states that have not yet been measured.

The following is organized into three sections. In Sec.~II, we provide a brief introduction to the theoretical framework used in DPS, along with the approach adopted in our main analysis. In Sec.~III, we present the method used to fit the experimental data,  extract the values of $\sigma_{\text{eff}}$ for different final states, and show our predictions. Finally, in Sec.~IV, we summarize our main findings and highlight their implications.

\section{Theoretical framework}
\label{Sec:formalism}

In this section, we present the theoretical formalism of double parton scattering on which our analysis is based. The inclusive DPS cross--section for hadrons $h$ and $h'$ in the collinear approximation, leading to final states $A$ and $B$, is given by:
\begin{align}
\sigma^{\text{DPS}}(AB) &=& \frac{m}{2} \sum_{i j; k' l'} \int  dx_1 dx_2 dx_1' dx_2'\ d^2 r \ \Gamma_{i j}^h(x_1, x_2; r) \hat{\sigma}_{i k'}^A (x_1, x_1') \hat{\sigma}_{j l'}^B (x_2, x_2') \Gamma_{k' l'}^{h'}(x_1', x_2'; r). 
\label{eq:collinear_factorization}
\end{align}
The total DPS cross--section is expressed as a convolution of double parton distribution functions (DPDFs) and parton cross sections.
The DPDF $\Gamma_{ij}^h$ gives the probability density of finding two partons of flavors $i$ and $j$ in hadron $h$, carrying longitudinal momentum fractions $x_1$ and $x_2$, separated by a transverse distance $r$, with an analogous description for $\Gamma_{k'l'}^{h'}$. These nonperturbative distributions depend on the internal structure of the hadrons. The perturbative component of the process arises from the parton cross sections $\hat{\sigma}_{ik',,jl'}^{A, B}$, which describe the hard scatterings between parton pairs $(i, k')$ and $(j, l')$. We take the renormalization and factorization scales to be the energy hard scales $\mu_A$ and $\mu_B$ of the processes; this dependence is implicit in the above formula.

We rewrite the DPDFs as the products of two single PDFs $f_{i,j}^{h}$ and the two--parton transverse distribution $F^h_{ij}$:
\begin{align}
    \Gamma_{i j}^h(x_1, x_2; r\ | \mu_A, \mu_B) \equiv 
    f_i^h(x_1 | \mu_A) f_j^h(x_2 | \mu_B)
    F^h_{ij}(x_1, x_2; r\ | \mu_A, \mu_B).
    \label{eq:DPDFfactorization}
\end{align}
Expressing DPDFs in terms of the well-established single PDFs is advantageous, as the latter are known with high precision, thanks to extensive constraints from experimental data. Our $F^h_{ij}$ depends on exactly the same variables as $\Gamma_{i j}^h(x_1, x_2; r\ | \mu_A, \mu_B)$; therefore, the above formula is not an approximation. 

Here, in our main analysis, we drop the dependence on parton kind or flavor. We keep the dependence on longitudinal momentum fractions; otherwise, Eq.~\ref{eq:collinear_factorization} reduces to the standard pocket formula (Eq.~\ref{eq:pocketformula}), and the effective cross section $\sigma_{\text{eff}}$ becomes the same for all final states. We are only interested in proton--proton and proton--antiproton collisions, so, by isospin symmetry, the antiproton structure can be obtained from the proton one.

By substituting Eq.~\ref{eq:DPDFfactorization} into Eq.~\ref{eq:collinear_factorization}, the DPS cross section can thus be expressed as:
\begin{align}
    \sigma^{\text{DPS}}(AB) & = \frac{m}{2} \int dx_1 dx_2 dx_1' dx_2'\ \Theta(x_1, x_2; x_1', x_2') 
    \sigma^A (x_1, x_1') \sigma^B(x_2, x_2') .
    \label{eq:factorizedcrosssection}
\end{align}
The $\Theta$ scale factor is given by:
\begin{align}
  \Theta(x_1, x_2; x_1', x_2'\, | \mu_A, \mu_B) = \int d^2 r \ F(x_1, x_2; r\, | \mu_A, \mu_B) F(x_1', x_2'; r\, | \mu_A, \mu_B).
   \label{eq:scalefactor}
\end{align}
The scale factor is a geometrical coefficient with the dimension of the inverse of a cross section, encapsulating information about parton transverse distribution. In Eq.~\ref{eq:factorizedcrosssection}, we also have the differential SPS cross section:
\begin{align}
    \sigma^{A} (x_{1}, x'_{1}|\mu_{A})
    = \frac{d \sigma^{\text{SPS}}(A)}{d x_{1} d x'_{1}} 
    = \sum_{ik'} f_{i}^{h}(x_{1} | \mu_{A}) f_{k'}^{h'}(x'_{1} | \mu_{A})
    \hat{\sigma}_{i k'}^{A} (x_{1}, x'_{1}\, | \mu_{A}).
    \label{eq:unintegratedcrosssection}
\end{align}
The analogous definition holds for the $B$ process as a function of $x_2$, $x'_2$, and $\mu_B$.

We are interested in the different measured values of $\sigma_{\text{eff}}$ for various $A$ and $B$. By using Eq.~\ref{eq:pocketformula} and Eq.~\ref{eq:factorizedcrosssection}, we define them to be: 
\begin{align} \label{eq:sigmaeff}
    \sigma_{\text{eff}}(AB) 
     =& \frac{\displaystyle  \int dx_1 dx'_1\, 
     \sigma^A (x_1, x_1')   
      \int dx_2 dx_2'\, 
     \sigma^B (x_2, x_2')\,  }
    {\displaystyle  \int dx_1 dx_2 dx_1' dx_2'\ \Theta(x_1, x_2; x_1', x_2')\ \sigma^A (x_1, x_1')   
     \sigma^B (x_2, x_2') }. 
\end{align}
We stress that $\sigma_{\text{eff}}(AB)$ is no longer universal; it now depends on the processes and their initial--state, as each SPS cross section weights the $\Theta$ scale factor differently. Consequently, by knowing $\sigma_{\text{eff}}(AB)$ for various final states, one can determine the $x$-behavior of the scale factor, thereby extracting information about the transverse distributions of the hadrons.

We adopt a 2D isotropic Gaussian parameterization for $F$, centered at $r = 0$, as is common in the literature, for example, Refs.~\cite{Sjostrand:1987su, Diehl:2014vaa, Frankfurt:2010ea}:
\begin{align}
    F(x_1, x_2; r\ | \mu_{AB}) = \frac{H(1-x_1-x_2)}{2\pi B(x_{12}, \mu_{AB})}\ \mathrm{exp}\left[-\frac{r^2}{2B(x_{12}, \mu_{AB})} \right].
    \label{eq:gaussian_hip}
\end{align}
In the variance $B(x,\mu)$, we adopt the geometric mean of the momentum fractions, $x_{12} := \sqrt{x_1 x_2}$, and of the hard scales, $\mu_{AB} := \sqrt{\mu_A \mu_B}$. The Heaviside function $H$ enforces the kinematic constraint $x_1 + x_2 \leq 1$, which becomes particularly relevant at forward rapidities. However, most of the $\sigma_\text{eff}$ data does not reach $x$ very close to 1. The Heaviside function gives the restriction $x_{12} \leq 0.5$.

After integrating over the transverse distance in Eq.~\ref{eq:scalefactor}, the resulting expression for the scale factor function is given by:
\begin{align}
 \Theta(x_1, x_2; x_1', x_2' | \mu_{AB}) 
    & = \frac{1}{2\pi} \frac{H(1-x_1-x_2) H(1-x'_1-x'_2)}{B(x_{12}, \mu_{AB}) + B(x'_{12}, \mu_{AB})}.
\end{align}
The $B(x, \mu)$ parameter provides crucial information about how the longitudinal momentum fraction and the hard scale affect the transverse double distribution, and its square root gives an estimate of the mean transverse distance between the partons in the pair. This is significant, as the closer the partons are in transverse space, the greater the likelihood of DPS occurrence.

\section{Results and predictions}
\label{Sec:prediction}

The final ingredient in our analysis is the shape of the Gaussian variance function $B(x, \mu)$. In particular, it indicates whether the two partons are, on average, close or far apart in transverse space. We use the following function with four parameters ($\beta$, $\gamma_1$, $\gamma_2$, and $\kappa$):
\begin{align}
   B(x, \mu) = \beta + \gamma_1 H(x_0 - x) \ln\left(\frac{x_0}{x}\right) + \gamma_2 H(x - x_v) + \kappa \ln\left(\frac{\mu}{\mu_0}\right).
   \label{eq:width}
\end{align}
This parameterization is inspired by Gribov diffusion for small $x$ and for large $\mu$~\cite{Gribov:1973jg, Brodsky:1994kf} and our previous work on valence or sea quarks~\cite{Huayra:2023gio}. 
The parameter $\gamma_1$ governs the low-$x$ behavior restricted by the Heaviside function to the region $x \leq x_0 = 0.001$, while $\gamma_2$ characterizes an additional separation for (valence) partons at momentum fractions larger than $x_v = 0.01$. The final term, proportional to $\kappa$, introduces a logarithmic dependence on the scale $\mu$, representing the scale evolution of the variance due to QCD dynamics. By dimensional considerations, we define the reference scale as $\mu_0 = 1$\,GeV.

\begin{table}
\def\arraystretch{1.4
}
    \centering
    \begin{tabular}{|c|c|c|c|c|} \hline
    Parameters & $\beta$ & $\gamma_1$ & $\gamma_2$ & $\kappa$  \\  \hline \hline
     Values (mb) & $0.067 \pm 0.068$ & $1.68 \pm 0.48$ & $0.85 \pm 0.16$ & $0.087 \pm 0.036$ \\ \hline
    \end{tabular}
    \caption{Global Gaussian variance parameters used to parameterize the parton pair distributions in transverse space. In our fit, we find an optimum $\chi_{\text{dof}}^2 = 29.6/(26-4) = 1.35$.}
    \label{tab:parametersx}
\end{table}

In order to determine the coefficients $\beta$, $\gamma_1$, $\gamma_2$, and $\kappa$, we fit Eq.~\ref{eq:sigmaeff} to the available experimental data~\cite{CDF:1993sbj, CDF:1997yfa, ATLAS:2013aph, D0:2014owy, D0:2014vql, CMS:2013huw, D0:2015rpo, CMS:2015wcf, ATLAS:2016rnd, LHCb:2015wvu, ATLAS:2016ydt, LHCb:2016wuo, Lansberg:2017chq, Lansberg:2019adr, CMS:2019jcb, CMS:2021lxi, CMS:2022pio, ALICE:2023lsn, Leontsinis:2022cyi, LHCb:2023qgu, LHCb:2023ybt}. The data comprise various final states, such as $J/\psi$, $\Upsilon$, jets, photons, and $Z$ and $W$ bosons, from the CDF, D0, ALICE, ATLAS, CMS, and LHCb experiments; thus, we perform a global fit.

\begin{figure*}[tb]
\centering
\includegraphics[width=\linewidth]{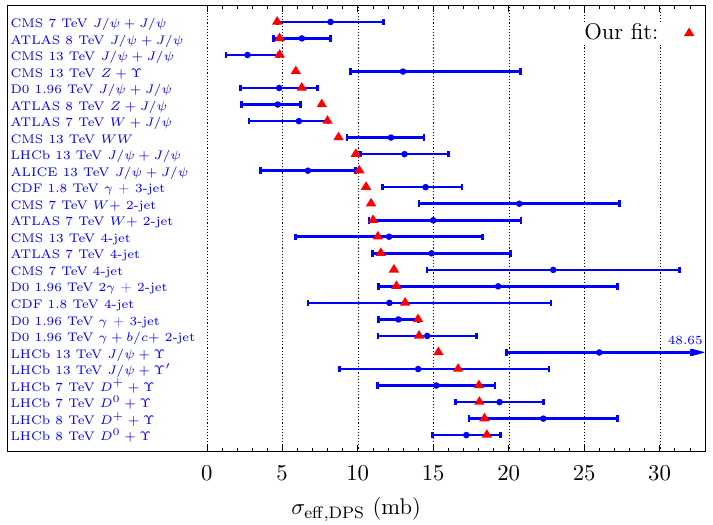}
\caption{Double parton scattering effective cross section $\sigma_{\text{eff}}(AB)$ measurements~\cite{CDF:1993sbj, CDF:1997yfa, ATLAS:2013aph, D0:2014owy, D0:2014vql, CMS:2013huw, D0:2015rpo, CMS:2015wcf, ATLAS:2016rnd, LHCb:2015wvu, ATLAS:2016ydt, LHCb:2016wuo, Lansberg:2017chq, Lansberg:2019adr, CMS:2019jcb, CMS:2021lxi, CMS:2022pio, ALICE:2023lsn, Leontsinis:2022cyi, LHCb:2023qgu, LHCb:2023ybt} compared to the results from our fitted model. 
}
\label{fig:results}
\end{figure*}

For Eq.~\ref{eq:sigmaeff} to be used in the fit, we need $\sigma_{ik'}^{A,B}$ as a function of $x_{1,2}$, $x'_{1,2}$ and $\mu_{A,B}$. We obtain these SPS cross sections from PYTHIA8.3~\cite{Bierlich:2022pfr} with the default NNPDF2.3 QCD+QED LO parton distribution functions~\cite{NNPDF:2017mvq}. The experimental kinematic cuts of all $A$ and $B$ processes are properly taken into account. We employ the default PYTHIA parameters based on the assumption that the generator is set to describe a wide range of high--energy processes across different experiments---aligning with the spirit of our global fit. The only modification made is the deactivation of multiparton interactions (MPI) since we are interested only in the SPS cross section. The $\mu_{A,B}$ hard scales are given by the default factorization scale used by PYTHIA. Considering all datapoints included in the fit, the largest contribution comes from partons in the $x$-interval $[10^{-3}, 10^{-1}]$, but with a sizable contribution from $x \approx 10^{-4}$ due to the forward rapidity measurements.

We perform the fit using the library Minuit2 \cite{James:2004xla} to minimize the reduced chi-squared accounting for experimental uncertainties. 
We obtain the following result:
\begin{align}
\chi^2_{\text{dof}} = 29.6/(26 - 4) = 1.35.
\end{align}
The parameters found in the fit can be seen in Tab.~\ref{tab:parametersx}. In Fig.~\ref{fig:results}, we present the fitted experimental data as blue points with uncertainty bars and the results of our fit as red triangles. There is good overall agreement between the experimental data and the calculated values.

\begin{figure}
    \centering
    \includegraphics[width=0.8\linewidth]{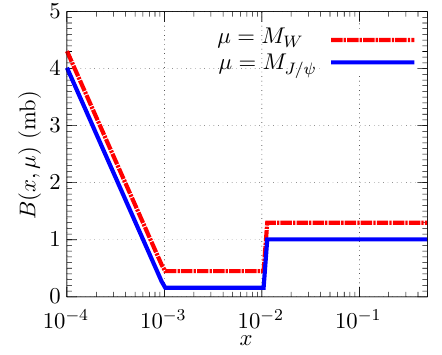}
    \caption{Parameterization of the variance function $B(x,\mu)$ with scales fixed by the masses $\mu_{AB} \equiv m_{J/\psi} = 3.096$\,GeV (blue line) and $\mu_{AB} \equiv m_{W} = 80.377$\,GeV (red line). At small and large values of the longitudinal momentum fraction $x$, the variance of the Gaussian distribution increases, indicating that the partons within the pair are more widely separated in the transverse plane.}
    \label{fig:B_function}
\end{figure}

The fitted function $B(x,\mu)$ is shown in Fig.~\ref{fig:B_function}. We see that, at small values of $x$, the Gaussian variance is large, meaning that partons are more widely separated in the transverse plane, which suppresses DPS. This is also seen in ultraperipheral collisions~\cite{Huayra:2019iun, Huayra:2020iib, Huayra:2021eve}. As $x$ grows, DPS becomes more frequent due to decreasing $B(x,\mu)$. In the interval $10^{-3} < x < 10^{-2}$, the Gaussian width is constant, yielding the smallest effective cross section possible: $\sigma_\text{eff}(J/\psi J/\psi) |_{10^{-3} < x < 10^{-2}} \approx 2$\,mb. Such a small cross section is compatible with existing models, e.g., Ref.~\cite{Gotsman:2019bol}. For $x > 10^{-2}$, in the valence--dominated range, $B(x,\mu)$ is also constant but $0.85$\,mb larger, with a sudden increase at $x_v = 10^{-2}$. We have attempted to fit a smoother $B(x,\mu)$, but the data prefer this step function.

When we do not include the final term in Eq.~\ref{eq:width}, i.e, the hard scale log term, we get a significant increase in the chi--square in our fit around $\Delta \chi^2 \sim$ 4.5. We have also performed a fit allowing the $x_0$ and $x_v$ parameters to vary, it did not improve the fit or change our results considerably.

We expect that our results would not change significantly if another Monte Carlo generator or PDF set were used, provided they are also designed to globally describe observables using LO parton cross sections. An important advantage of the expression in Eq.~\ref{eq:sigmaeff}, and of the measurements of the effective cross sections, is that uncertainties present in the individual SPS cross sections cancel in the ratio. This is a key reason why we chose to fit $\sigma_{\text{eff}}$ directly, rather than the DPS cross section itself.

To evaluate the statistical significance of our results and the correlations found, we also fit the fully--factorized pocket formula as the null hypothesis. We obtain a reduced chi-squared of $(\chi^2_{\text{dof}})_{\text{null}} = 94.9 / (26 - 1) = 3.80$ and an effective cross section of $\sigma_{\text{eff}} = 10.35 \pm 0.54$\,mb ($B = 0.824$\,mb), common to all data points. Therefore, the null hypothesis is rejected with a statistical significance greater than $6\sigma$. Moreover, our model is statistically superior to one with constant $\sigma_\text{eff}$, providing an improvement in fit that is significant at the 7.6$\sigma$ level.

 We also consider our previous hypothesis~\cite{Huayra:2023gio} of (sea and valence) parton--kind correlations (without $x$-dependence and with only two free parameters), originally fitted to only 18 datapoints. Redoing the fit with the current dataset of 26 points yields the reduced chi-squared value of $\chi^2_{\text{dof}} = 91.6/(26 - 2) = 3.82$. This indicates that introducing an explicit $x$-dependence in the double transverse distributions results in a better model than the previously considered parton--kind correlations. We understand that any parton--kind dependent correlations will be captured, in good part, by the $x$-dependence of $B(x,\mu)$, since valence partons typically have larger $x$.

\begin{figure}
    \centering
    \includegraphics[width=\linewidth]{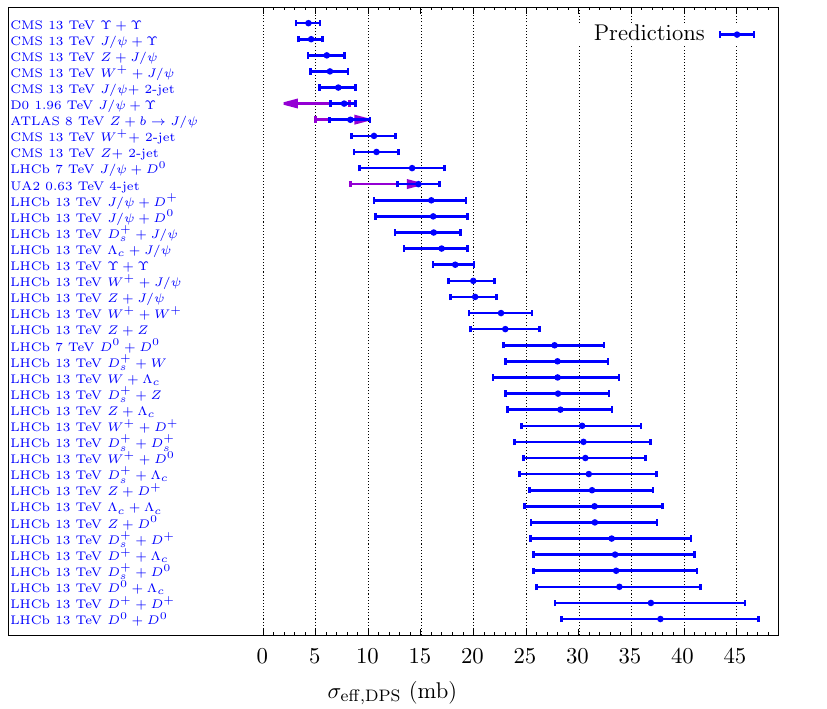}
    \caption{Predictions of $\sigma_{\text{eff, DPS}}$ in $pp$ collisions for various final states from the global fit of our model. The kinematic cuts are the same as those used by the experiments in DPS analyses (when available) or in corresponding SPS analyses. The purple arrows indicate upper or lower experimental limits for D0 Ref.~\cite{Shao:2016wor}, ATLAS Ref.~\cite{Lansberg:2016muq}, and UA2 Ref.~\cite{Alitti1991ASO}.}
    \label{fig:predictions}
\end{figure}

With our model and fitted parameters, predictions for DPS effective cross sections not yet measured are a straightforward application. We present some in Fig.~\ref{fig:predictions}, for a variety of final states accessible at the LHC, such as combinations of $D^{\pm, 0}$, $D^{\pm}_s$, $W$, $Z$, etc, at central and forward experiments. As such, future measurements of these processes, particularly in the forward region or in channels involving electroweak bosons at central rapidities, can serve as tests and validations of our results. In Fig.~\ref{fig:predictions}, we have used the same kinematic cuts as in Fig.~\ref{fig:results} when we have the same SPS observable and experiment. We also used the cuts from Ref.~\cite{LHCb:2012aiv} for $\Lambda_c$ and $D_s^+$ at LHCb and Ref.~\cite{LHCb:2015mad} for $W$ and $Z$ boson at LHCb.

We remark that the $D^0 + D^0$ production at the LHC gets a predicted effective cross section of about 40\,mb, significantly larger than the fitted value of about 10\,mb of the $J/\psi+J/\psi$ production. In the case of $D^0$ production, the target side $x$ value is concentrated in the range from $10^{-6}$ to $10^{-4}$, while in the case of $J/\psi$ production, it is mostly concentrated in the range from $10^{-5}$ to $3 \cdot 10^{-3}$, i.e, larger $x$. At the matrix element, the $D^0$ requires almost always the $c \bar{c}$ creation, while $J/\psi$ requires the pair plus another parton by color considerations. This is why the latter needs larger $x$ and will produce a smaller effective cross section.

\section{Conclusions}
\label{Sec:conclusions}

This work studies double parton scattering (DPS) in proton--proton collisions. This type of observable is sensitive to correlations between the initial parton pair, unlike in the SPS case. The fully factorized approach provides the simplest pocket formula, where most correlations are neglected and the effective cross section $\sigma_{\text{eff}}$ is the only remaining parameter. This observable has been measured for a variety of final states, with values ranging from 2.7 to 26\,mb.

In our analysis, we considered the possibility that the double parton distribution is the product of single parton distributions and a two--parton transverse distribution that depends on the parton longitudinal momentum fractions and the hard scales of the processes. This led to a generalized pocket formula, Eq.~\ref{eq:factorizedcrosssection}, which accounts for correlations depending on the specific final state considered. 

To model the transverse distance between parton pairs, we adopted a Gaussian profile with a variance given by a function $B(x, \mu)$. This framework enables us to compute the quantity measured in experiments, $\sigma_{\text{eff}}(AB)$, using PYTHIA to obtain the SPS cross sections. One caveat is that, of course, other types of correlations between partons, aside from transverse ones, are possible, and in our framework, they can be reinterpreted as transverse correlations.

We analyzed 26 measurements of $\sigma_{\text{eff}}$ from two hadron colliders and six detectors, covering different center--of--mass energies and kinematic cuts. These data were used in a global fit of $B(x, \mu)$ modeled with four parameters. We obtained a goodness-of-fit of $\chi^2_{\text{dof}} = 29.6/(26 - 4) = 1.35$. The extracted values of $\sigma_{\text{eff}}(AB)$ show good agreement with experimental data, demonstrating that including $x$- and $\mu$-dependence between parton populations significantly improves the description of DPS.

Our results represent a significant improvement over the naive approach that assumes a single universal value for $\sigma_{\text{eff}}$. The improved $\chi^2_{\text{dof}}$ compared to our previous result~\cite{Huayra:2023gio} also indicates that the inclusion of $x$-dependence in the double transverse distributions plays a more relevant role than the previously assumed (sea or valence) parton--kind correlations.

We observed the following behavior in the mean transverse distance between partons: they are farther apart at small $x$, where the partons are more spatially spread out; become closer at intermediate $x$, when correlations concentrate them; and again more distant at large $x$, the characteristic valence-like behavior. The larger $\sigma_{\text{eff}}$ values observed by LHCb can be understood from this pattern, as forward rapidity configurations typically involve one small--$x$ and one large--$x$ parton. Additionally, we found that the parton separation increases logarithmically with the hard scale of the process.

Finally, we present quantitative predictions for $\sigma_{\text{eff}}(AB)$ in specific channels not yet measured. These results can be tested in future measurements and can guide a more systematic exploration of DPS. Moreover, our model enables the generation of double parton distributions from the well--established single parton distributions. The broad scope of our analyses represents a significant advance in the understanding of the proton structure.

\section*{Acknowledgments}

This work was supported by FAPESC, INCT-FNA (464898/2014-5), and National Council for Scientific and Technological Development – CNPq (Brazil) for JVCL, EH, and EGdO. This study was financed in part by the Coordenação de Aperfeiçoamento de Pessoal de Nível Superior -- Brasil (CAPES) -- Finance Code 001. 



\end{document}